\documentclass[twocolumn,showpacs,prb]{revtex4}
\usepackage{graphics}
\usepackage{epsfig}
\usepackage{dcolumn}
\usepackage{amsmath}
\usepackage{color}

\newcommand{\beq}{\begin{equation}}
\newcommand{\eeq}{\end{equation}}
\newcommand{\bea}{\begin{eqnarray}}
\newcommand{\eea}{\end{eqnarray}}
\newcommand{\bwt}{\begin{widetext}}
\newcommand{\ewt}{\end{widetext}}

\begin{document}

\title{Necessary and sufficient condition for longitudinal magnetoresistance}
\date{\today}
\author{H. K. Pal and D. L. Maslov}

\begin{abstract}
Since the Lorentz force 
is perpendicular to the magnetic field, it should not affect the motion of a charge along the field. This argument seems to imply absence of longitudinal magnetoresistance (LMR) which is, however, 
observed in many materials and reproduced by standard semiclassical transport theory applied to particular metals.  We derive a necessary and sufficient condition on the shape of the Fermi surface for non-zero LMR. 
Although an anisotropic spectrum is a pre-requisite for LMR, not all types of anisotropy can give rise to the effect: a spectrum
should not be separable in any sense. More precisely, the combination $k_{\rho}v_{\phi}/v_{\rho}$, where $k_\rho$ is the radial component of the momentum in a cylindrical system with the $z$ axis along the magnetic field and $v_{\rho} (v_{\phi}$) is the radial (tangential) component of the velocity, should depend on the momentum along the field. 
For some lattice types, this condition is satisfied already at the level of nearest-neighbor hopping;  for others, the required non-separabality occurs only if next-to-nearest-neighbor hopping is taken into account. 
\end{abstract}

\affiliation{
Department
of Physics, University of Florida,
Gainesville, FL 32611-8440, USA}

\maketitle
\section{Introduction}

Magnetoresistance, i.e., a change in the resistance due to a magnetic
field, can be distinguished into two types depending on the mutual
orientation of the current and the magnetic field: transverse (TMR) and
longitudinal (LMR). Although a change in the transverse
resistance due to a magnetic field is natural because electrons experience
Lorentz force in that direction, the very existence of LMR is somewhat
surprising, at least at first glance. Indeed, since Lorentz force is
perpendicular to the field, one does not expect the motion of electrons along the
field to be affected. A weak point of this argument is that it applies,
strictly speaking, only to free electrons but not to electrons in metals.
Moreover, LMR is absent in a more 
realistic (yet still incomplete) 
``damped Bloch electrons model" (DBEM), in which a phenomenological damping term is
introduced into the semiclassical equations of motion for an arbitrary
electron spectrum. \cite{Kittel,Omar,Marder_1} However, we will argue in this paper that
the ``damped Bloch electrons model'' 
is not equivalent to
the
Boltzmann equation, which provides the only complete semiclassical description
of semiclassical dynamics of electrons in solids in the presence of scattering.
Therefore, absence of LMR in DBEM does not imply its absence in reality.

Experimentally, LMR has been observed in many materials. \cite{pip1,pip2}
Theoretically, a general solution of the Boltzmann equation in the magnetic
field does not exclude LMR;\cite{phy} calculations performed for particular
metals, e.g., copper, do yield finite LMR.~\cite{pip1,pip2} However, it is not clear from
this general solution which symmetries must be broken, i.e., how anisotropic
the electron spectrum should be for LMR to occur. It is probably why LMR is
sometimes
viewed as a kind of surprise.\cite{hus,lai} In addition to anisotropic spectrum, several
more special models have been invoked to explain LMR. It was shown, for
example, that LMR can arise due to anisotropic scattering, \cite{son1}
macroscopic inhomegeneities, \cite{stroud} including
barrier inhomogeneities in superlattices, \cite{lai}
as well as due to a
modification of the density of states by the magnetic field in the
ultra-quantum regime, when all but the lowest Landau levels are depopulated. 
\cite{arg} Whereas observed LMR in many cases is likely to be caused by these
more evolved mechanisms, it is still necessary to explore whether LMR can
arise simply due to anisotropy of the Fermi surface (FS) and to formulate a
minimal condition for LMR to occur.

Magnetotransport in non-quantizing fields is described by the Boltzmann
equation which gives the conductivity tensor. To find magnetoresistance, one
inverts this tensor. It is well known that for any isotropic spectrum the
magnetic field dependences of the diagonal and off-diagonal conductivities
cancel out, so that both TMR and LMR are absent. While TMR can be made
finite by either invoking any kind of anisotropy of the Fermi surface or
introducing a multiband picture while keeping the spectrum isotropic, the
story with LMR is not so simple. As is shown in this paper, not all types of anisotropy give rise to LMR, e.g., deforming a spherical Fermi surface
into an ellipsoidal one is not enough. We derive the necessary and
sufficient condition the spectrum must satisfy for LMR to occur and discuss 
the implications of this condition for several types of bandstructure.
For example, metals with 
face-centered cubic (FCC) and body-centered cubic (BCC) lattices satisfy the necessary and sufficient condition even if only nearest-neighbor hopping is taken into account, whereas a simple cubic (SC) lattice has LMR only due to hopping between next-to nearest neighbors.The same is true for layered structures, such as hexagonal planes stacked on top of each other, where one has to include out-of-plane next-nearest-neighbor interactions to see the effect.

The rest of the paper is organized as follows. In Sec.~\ref{sec2} we show
that
LMR is absent in 
DBEM and 
analyze the differences 
between this and Boltzmann-equation approach. In Sec.~\ref{sec3},
we derive the necessary and sufficient condition for LMR in the
Boltzmann-equation formalism 
and discuss the implications of this condition. 
As a particular example, we consider the case of Bernal-stacked
graphite in Sec.~\ref{sec5}. In graphite, the necessary and sufficient
condition is satisfied due to trigonal warping of the Fermi surface. We
find, however, that strong non-parabolic LMR observed in highly oriented
pyrolytic graphite (HOPG) samples \cite{bra} cannot be accounted for LMR 
arising simply from
anisotropy of the Fermi surface. Our conclusions
are given in Sec.~\ref{sec6}.

\section{\label{sec2} Semiclassical equations of motion}

The effect of weak electric and magnetic fields 
on electrons in solids can be described by the
semiclassical equations of motion \cite{ash} 
\begin{mathletters}
\begin{eqnarray}
\mathbf{v} &=&\frac{\partial \varepsilon _{\mathbf{k}}}{\partial \mathbf{k}},
\label{kdot_a} \\
\frac{d\mathbf{k}}{dt} &=&e(\mathbf{E}+\mathbf{v}\times \mathbf{B}),
\label{eq:kdot}
\end{eqnarray}
where $e$ is the electron charge and we set $\hbar =1$. We neglect here the
anomalous terms in the velocity which, even if present, are small in weak
magnetic fields.~\cite{sundaram99,Marder_2} In the absence of scattering,
Eqs.~(\ref{kdot_a}) and (\ref{eq:kdot}) are valid for an arbitrary spectrum $%
\varepsilon (\mathbf{k})$ and provide an invaluable tool for analyzing
collisionless dynamics of electrons in solids. To account for scattering of
electrons by impurities, phonons, etc., it is customary to replace Eqs.~(\ref{kdot_a}) and (\ref{eq:kdot}) by a phenomenological "damped Bloch electron model" (DBEM)
with a damping term $-\mathbf{k}/\tau $ inserted into the right-hand side of Eq.~(\ref{eq:kdot}). 
\cite{Kittel,Omar,Marder_1}  In steady-state, DBEM reduces to 
\end{mathletters}
\begin{equation}
\frac{\mathbf{k}}{\tau }=e(\mathbf{E}+\mathbf{v}\times \mathbf{B}).
\label{eq:kdoteq}
\end{equation}
We are now going to show that this approach eliminates LMR not only for an
isotropic but also for an arbitrary spectrum. To find LMR, we assume that the current $%
\mathbf{j}=en_{c}\mathbf{v}$, where $n_c$ is the number density of conduction
electrons, is along $\mathbf{B}$ chosen as the $z$-axis. Then, 
\begin{mathletters}
\begin{eqnarray}
v_{z}(k_{x},k_{y},k_{z}) &=&\frac{j_{z}}{n_{c}e}, \label{eq13}\\
v_{x}(k_{x},k_{y},k_{z}) &=&0, \\
v_{y}(k_{x},k_{y},k_{z}) &=&0.
\end{eqnarray}
Furthermore, the equation of motion ~(\ref{eq:kdoteq}) for the $k_{z}$
component gives 
\end{mathletters}
\begin{equation}
k_{z}=e\tau E_{z}.  \label{eq4}
\end{equation}
The set of four equations (\ref{eq13}-\ref{eq4}) defines an inhomogeneous
system for four unknowns: $k_{x},k_{y},k_{z}$, and $E_{z}$. In general, such
a system has a unique solution. Therefore, $E_{z}$ can be found as a
function of $j_{z}$ using only Eqs.~ (\ref{eq13}-\ref{eq4}). Since none of these equations involve the magnetic
field, the longitudinal resistivity $\rho _{zz}=j_{z}/E_{z}$ does not
depend on $B$ either, which implies that LMR is absent for an arbitrary
spectrum. On the other hand, components $E_{x}$ and $E_{y}$ have to be found
from the equations of motion for $k_{x}$ and $k_{y}$ which do involve $B$,
and hence TMR is not zero for an arbitrary spectrum.

If the above conclusion were correct, it would be in variance with
experimental observations. As we will show shortly, non-zero LMR can be
understood only by using the Boltzmann equation 
\begin{equation}
\frac{df_{\mathbf{k}}}{dt}=\frac{\partial f_{\mathbf{k}}}{\partial t}+%
\mathbf{v}\cdot \frac{\partial f_{\mathbf{k}}}{\partial \mathbf{r}}+\mathbf{%
\dot{k}}\cdot \frac{\partial f_{\mathbf{k}}}{\partial \mathbf{k}}=I_{\mathrm{%
c}}\left[ f_{\mathbf{k}}\right] ,  \label{be}
\end{equation}
where $I_{\mathrm{c}}$ denotes the collision integral. 
Although the Boltzmann equation is a semi-classical description just like
the previous method, there is some conceptual difference between the two
approaches. The problem is that while the equations of motions in the
absence of scattering can be derived from the Schroedinger equation, the
DBEM does not follow from any microscopic
approach. Indeed, the momentum $\mathbf{k}$ in the absence of scattering
still has the meaning of the quantum number parameterizing the Bloch state $%
\psi _{\mathbf{k}}\left( \mathbf{r}\right) $. Hence a (slow) evolution of $%
\mathbf{k}$ with time in the presence of electric and magnetic fields
describes the evolution of $\psi _{\mathbf{k}}\left( \mathbf{r}\right) $. In
the presence of scattering, e.g., by disorder, $\psi _{\mathbf{k}}\left( 
\mathbf{r}\right) $ becomes a random quantity whose average over disorder
realizations does not have a particular meaning. Therefore, it is not
surprising that an \emph{ad hoc} insertion of the damping term into the
equation of motion does not capture essential physics. The shortcomings of
this procedure become obvious even in the absence of the magnetic field. For
example, Eq. (\ref{eq:kdoteq}) predicts that $\mathbf{k}$ is always parallel
to $\mathbf{E}$ if $\mathbf{B=0.}$ However, the \emph{average }momentum per
electron $\langle \mathbf{k\rangle =}\int d^{D}k\mathbf{k}f_{\mathbf{k}}/n,$
where $n$ is the total number density, is not parallel to $\mathbf{E}$ for a
lattice of sufficiently low symmetry. Indeed, solving Eq. (\ref{be}) in the
relaxation-time approximation at zero temperature yields $\langle \mathbf{%
k\rangle }=e\tau \oint dS_{F}(\mathbf{v\cdot E)k/}v_{F}\left( \mathbf{k}%
\right) $, where $dS_{F}$ is the element of the Fermi surface and $%
v_{F}\left( \mathbf{k}\right) $ is the magnitude of the electron velocity at
a given point on this surface. For a generic Fermi surface, $\langle \mathbf{%
k\rangle }$ and $\mathbf{E}$ are not parallel. Also, the conductivity given
by the 
DBEM as $\sigma ^{\prime
}=e^{2}v_{F}^{2}\tau \nu (\varepsilon _{F})/3$, where $\nu (\varepsilon )$
is the density of states, coincides with the result of the Boltzmann
equation $\sigma _{\alpha \beta }=e^{2}\int dS_{F}v_{\alpha }v_{\beta }\tau
/v_{F}(\mathbf{k})$ only for an isotropic spectrum.

\section{\label{sec3} Minimal conditions for 
longitudinal magnetoresistance
}

\subsection{Necessary condition}

Having dealt with the inconsistencies of the ``damped Bloch electrons
model'', we now return to the original problem of finding the minimum
requirement for non-zero LMR for an arbitrary spectrum $\varepsilon
=\varepsilon (\mathbf{k})$. In the linear-response regime, one can rewrite
Eq.~(\ref{be}) for the non-equilibrium part of the distribution function $g(%
\mathbf{k})=f_{\mathbf{k}}-f_{\mathbf{k}}^{0}$ as 
\begin{eqnarray}
\left( 1+{\hat{\Omega}}\right) g(\mathbf{k}) &\equiv &\left[ 1+e\tau (%
\mathbf{v}\times \mathbf{B})\cdot \frac{\partial }{\partial \mathbf{k}}%
\right] g(\mathbf{k})  \notag \\
&=&-e\mathbf{E}\cdot \mathbf{v}\frac{\partial f_{\mathbf{k}}^{0}}{\partial
\varepsilon _{\mathbf{k}}},  \label{boltzmannproper}
\end{eqnarray}
where we have also adopted the relaxation-time approximation (which is exact
for 
isotropic
impurity scattering). Since we are interested in the minimal condition, we allow $\tau $ to depend
only on $\varepsilon $ but not on the direction of $\mathbf{k}$ and assume
that all components of $\mathbf{k}$ relax at the same rate, i.e., that $%
1/\tau $ is a scalar rather than a tensor. We will come back to this point later in the paper. 
For $\mathbf{B}||{\hat{z}}$, 
\begin{equation}
\hat{\Omega}=\tau e(\mathbf{v}\times \mathbf{B})\cdot \frac{\partial }{%
\partial \mathbf{k}}=e\tau B\left( v_{y}\frac{\partial }{\partial k_{x}}%
-v_{x}\frac{\partial }{\partial k_{y}}\right) .  \label{omega}
\end{equation}
Following the Zener-Jones method, \cite{zim} we express $g(\mathbf{k})$ via
an infinite series in the operator $\hat{\Omega}$: 
\begin{eqnarray}
g(\mathbf{k}) &=&(1+\hat{\Omega})^{-1}\left( -\tau e\mathbf{E}\cdot \mathbf{v%
}\frac{\partial f_{\mathbf{k}}^{0}}{\partial \varepsilon _{\mathbf{k}}}%
\right)   \notag \\
&=&\sum_{n=0}^{\infty }(-\hat{\Omega})^{n}\left( -\tau e\mathbf{E}\cdot 
\mathbf{v}\frac{\partial f_{\mathbf{k}}^{0}}{\partial \varepsilon _{\mathbf{k%
}}}\right) .  \label{eq:omega}
\end{eqnarray}
Note that the operator $\hat{\Omega}$ always yields zero when it acts on any
function that depends on $\varepsilon _{\mathbf{k}}$ only. Hence, in Eq. (%
\ref{eq:omega}), $\hat{\Omega}$ acts only on $\mathbf{v}$. Substituting Eq. (%
\ref{eq:omega}) into the current $\mathbf{j=}2e\int d^{3}k\mathbf{v}g\left( 
\mathbf{k}\right) /\left( 2\pi \right) ^{3}$, we find the conductivity as 
\begin{equation}
\sigma _{\alpha \beta }=2e^{2}\tau \int \frac{d^{3}k}{(2\pi )^{3}}\left( -%
\frac{\partial f_{\mathbf{k}}^{0}}{\partial \varepsilon _{\mathbf{k}}}%
\right) v_{\alpha }\sum_{n=0}^{\infty }\left( -\hat{\Omega}\right)
^{n}v_{\beta }.  \label{current}
\end{equation}
In the LMR geometry, $\mathbf{E}||\mathbf{B}||{\hat{z}}$. If $\hat{\Omega}%
v_{z}%
=0$, all but the $n=0$ term in Eq.~(\ref{current}) are equal to zero.
Therefore, a \emph{necessary} condition
for $\sigma _{zz}$ to depend on the
magnetic field is 
\begin{equation}
{\hat{\Omega}}v_{z}\neq 0.  \label{omegacondition}
\end{equation}
Rewriting ${\hat{\Omega}}$ in cylindrical coordinates, the condition (\ref
{omegacondition}) can be re-expressed as: 
\begin{equation}
\left( \frac{\partial \varepsilon }{\partial \phi }\frac{\partial }{\partial
k_{\rho }}-\frac{\partial \varepsilon }{\partial k_{\rho }}\frac{\partial }{%
\partial \phi }\right) v_{z}\neq 0,
\end{equation}
or 
\begin{equation}
\frac{\partial }{\partial k_{z}}\left( \frac{\partial \varepsilon /\partial
\phi }{\partial \varepsilon /\partial k_{\rho }}\right) \neq 0.
\label{condition}
\end{equation}
On the other hand, Eq.~(\ref{condition}) is \emph{not} a sufficient condition
because even if ${\hat{\Omega}}^{n}v_{z}\neq 0$ for the $n^{\mathrm{th}}$
term in the series, the contribution of this term to $\sigma _{zz}$ may
vanish upon integrating over the Fermi surface. For example, since $\sigma
_{zz}$ must be an even function of $B$, all odd terms in the series must
vanish.

Equation (\ref{condition}) implies that the minimum condition on the
spectrum is that the ratio of $\partial \varepsilon /\partial \phi $ and $%
\partial \varepsilon /\partial k_{\rho }$ (equal to $k_{\rho }v_{\phi
}/v_{\rho }$) 
must depend on $k_{z}$. 
Geometrically, this means that the angle between the component of velocity perpendicular to the field and the radial direction at a given point on the Fermi surface must vary with $k_z$. It can be easily seen that if the spectrum does not depend
on $\phi $, condition (\ref{condition}) is trivially violated and there is
no LMR. Therefore, angular anisotropy of the FS about the magnetic-field
direction is a prerequisite. However, anisotropy must be of a special kind.
For example, spectra such as $\varepsilon _{\mathbf{k}}=\varepsilon
_{1}(k_{\rho },\phi )+\varepsilon _{2}(k_{z})$ and $\varepsilon _{\mathbf{k}%
}=\varepsilon _{1}(k_{\rho },\phi )\varepsilon _{2}(k_{z})$, which are
arbitrarily anisotropic in the $\phi $ direction but separable in $k_{z}$,
violate condition (\ref{condition}) and thus do not lead to LMR. 
As an 
example, let us consider 
a SC lattice with lattice parameter $a$.  In the tight-binding model  with nearest-neighbor hopping (parameterized by coupling $t_1$), the energy spectrum is given by 
$\varepsilon_{\mathbf{k}}=-2t_1\left[\mathrm{cos}(k_x a)+\mathrm{cos}(k_y a)+\mathrm{cos}(k_z a)\right]$ which, being separable in all three coordinates, clearly violates the LMR condition.  If next-to-nearest-neighbor hopping (parameterized by coupling $t_2$) is taken into account, additional terms $-4t_2[\mathrm{cos}(k_x a)\mathrm{cos}(k_y a)+\mathrm{cos}(k_y a)\mathrm{cos}(k_z a)+\mathrm{cos}(k_z a)\mathrm{cos}(k_x a)]$ occur in the spectrum, which no more violates the LMR condition. Thus, the effect comes only from next-to-nearest-neighbor hopping for an SC lattice. On the other hand, an FCC lattice satisfies the condition already at the nearest-neighbor level because the spectrum in this case  $\varepsilon_{\mathbf{k}}=-4t_1[\mathrm{cos}(k_x a/2)\mathrm{cos}(k_y a/2)+\mathrm{cos}(k_y a/2)\mathrm{cos}(k_z a/2)+\mathrm{cos}(k_z a/2)\mathrm{cos}(k_x a/2)]$ is non-separable; the same is true for a BCC lattice.
On the other hand, layered, e.g, hexagonal, structures, 
will require coupling between an atom located in one plane and another atom in the adjacent plane but situated obliquely from the former, if the magnetic field
is perpendicular to the planes (more on this later for the specific case of graphite).

\begin{figure}[t]
\includegraphics[angle=0,width=0.4\textwidth]{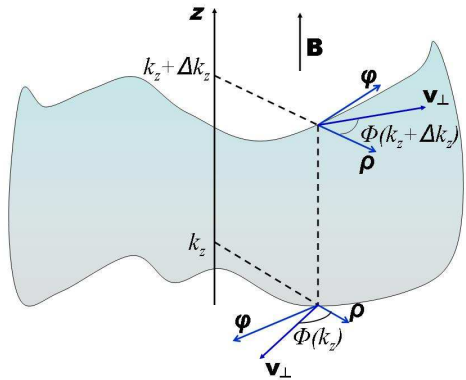}
\caption{Geometric interpretation of the necessary condition for
longitudinal magnetoresistance. Here,
$\mathbf{v_{\bot}}$
 is the component of the electron velocity perpendicular to the field.}
\label{fig1}
\end{figure}

A quantity measured in a typical experiment is not the conductivity but 
the resistivity. Generally speaking, the dependence of the conductivity on
the magnetic field does not automatically imply a dependence of the resistivity
on the field--a well known case is the isotropic spectrum, when the
(transverse) diagonal components of the conductivity depend on $B$ but the
diagonal components of the resistivity do not. 
It is necessary, therefore, to make sure that Eq.~(\ref{condition}) is not
only a necessary condition for longitudinal magnetoconductance but also for
magnetoresistance. 
It is difficult to prove that non-zero magnetoconductance implies non-zero LMR
for an arbitrary spectrum. 
To proceed further, we relax a condition on the energy spectrum, 
assuming that $B$ is perpendicular to the plane of symmetry, i.e., that $%
\varepsilon (k_{x},k_{y},k_{z})=\varepsilon (k_{x},k_{y},-k_{z})$. This
constraint is stronger than that imposed by 
time reversal symmetry (in the absence of the spin-orbit interaction and
magnetic structure), i.e., 
In this case, $v_{z}$ is odd while $v_{x}$ and $v_{y}$ are even in $k_{z}$, 
and the off-diagonal components $\sigma _{\alpha z}$ ($\alpha\neq z$) vanish both in  zero
and finite magnetic fields. For example, all terms in the expression for $%
\sigma _{xz}$ 
vanish
upon integration over $k_{z}$: 
\begin{eqnarray}
&&\sigma _{xz}=2e^{2}\tau \sum_{n=0}^{\infty }(-e\tau B)^{n}  \notag \\
&&\times \int \frac{d^{3}k}{(2\pi )^{3}}v_{x}\left( v_{y}\frac{\partial }{%
\partial k_{x}}-v_{x}\frac{\partial }{\partial k_{y}}\right) ^{n}v_{z}\frac{%
\partial f_{\mathbf{k}}^{0}}{\partial \varepsilon _{\mathbf{k}}}=0.
\end{eqnarray}
By the Onsager principle, $\sigma _{z\alpha }=$ as well. 
Therefore, the matrix of $\sigma _{\alpha \beta }$ is block-diagonal and $%
\rho _{zz}=1/\sigma _{zz}$. 
Thus Eq.~(\ref{condition}) is a \emph{necessary} condition for non-zero LMR
as well, provided that the spectrum is symmetric on inversion of $k_{z}$.

\subsection{Sufficient condition}

The condition presented in Eq.~(\ref{condition}) is only a necessary
condition for LMR, as the integral in Eq.~(\ref{current}) may still vanish
due to some symmetry even if the integrand satisfies Eq.~(\ref{condition}). To
formulate a sufficient condition, we approach the problem from the
strong-magnetic--field limit. In this limit, it is convenient to use the
method of Lifshitz, Azbel' and Kaganov, \cite{alk,abr,phy} in which the $k$%
-space is mapped onto a space defined by the set of variables $\varepsilon
\equiv \varepsilon _{\mathbf{k}},k_{z}$ and $t_{1}$, where   $t_{1}$, defined by
the equation 
\begin{equation}
\frac{d\mathbf{k}}{dt_{1}}=e\mathbf{v\times B,}  \label{time_1}
\end{equation}
is the time spent by an electron on the orbit in the $k$-space in the presence
of the magnetic field only. Accordingly, the integration measure is
transformed as 
\begin{equation}
\int \int \int dk_{x}dk_{x}dk_{z}=eB\int \int \int dt_{1}d\varepsilon dk_{z}.
\label{jac}
\end{equation}
The non-equilibrium correction to the distribution function can be written
as 
\begin{equation}
g=e\frac{\partial f^{0}}{\partial \varepsilon }\mathbf{E\cdot s,}
\end{equation}
where $\mathbf{s}$ satisfies 
\begin{equation}
\frac{\partial \mathbf{s}}{\partial t_{1}}=I_{\mathrm{c}}\left[ \mathbf{s}%
\right] +\mathbf{v.}
\end{equation}
Adopting the relaxation-time approximation for $I_{\mathrm{c}}$ and keeping
only the leading term in $1/B$ , it is easy to see that \cite{phy}
\begin{equation}
s_{z}=\tau \langle v_{z}\rangle ,
\end{equation}
where $\langle v_{z}\rangle =\frac{1}{T}\int v_{z}dt_{1}$ with $T$ being
either the period of an orbit (for closed orbits) or the time over which an
orbit reaches the boundary of the Brillouin zone (for open orbits). The $%
\sigma _{zz}$ component of the conductivity tensor in this limit is then equal to 
\begin{eqnarray}
\sigma _{zz}\left( \infty \right)  &=&\frac{2e^{2}\tau }{(2\pi )^{3}}eB\int
\int \int d\varepsilon dk_{z}dt_{1}v_{z}\langle v_{z}\rangle \left( -\frac{%
\partial f^{0}}{\partial \varepsilon }\right)   \notag \\
&=&\frac{2e^{2}\tau }{(2\pi )^{3}}\oint d\ell \int dk_{z}\int \frac{%
d\varepsilon }{v_{\perp }}v_{z}\langle v_{z}\rangle \left( -\frac{\partial
f^{0}}{\partial \varepsilon }\right),\nonumber\\  \label{sufcon}
\end{eqnarray}
where $d\ell $ is a line element along the orbit and $v_{\perp }=\sqrt{%
v_{x}^{2}+v_{y}^{2}.}$ Obviously, $\sigma _{zz}\left( \infty \right) $ does
not depend on $B.$ On the other hand, the zero-field value of $\sigma _{zz}$
is 
\begin{equation}
\sigma _{zz}\left( 0\right) =\frac{2e^{2}\tau }{(2\pi )^{3}}\int
d^{3}kv_{z}^{2}\left( -\frac{\partial f^{0}}{\partial \varepsilon }\right) .
\label{zerocon}
\end{equation}
Pippard\cite{pip2} suggested that the ratio 
$\sigma_{zz}(\infty)/\sigma_{zz}(0)$
may be used to get information about the scattering mechanisms on the FS. 
We, however, use 
this ratio to construct 
a sufficient condition for LMR. 
Keeping the same constraint on the energy spectrum 
$\varepsilon (k_{x},k_{y},k_{z})=\varepsilon (k_{x},k_{y},-k_{z})$  so that $\rho_{zz}=1/\sigma_{zz}$,
the \emph{sufficient} condition for LMR can now be formulated as follows: if $%
\sigma _{zz}(\infty )\neq \sigma _{zz}(0)$, we have non-zero LMR. It is only
a sufficient condition because, even if it is violated, LMR can still exist.
Indeed, even if asymptotic limits of the function $\sigma _{zz}\left(
B\right) $ coincide, it is not necessarily a constant. To formulate the
sufficient condition in more transparent terms, we use the following trick.
The integration measure in the expression (\ref{zerocon}) for the zero-field
conductivity can formally be re-written in terms of variables $\varepsilon
,k_{z}$ and $t_{1}$, as specified by transformation (\ref{jac}). Since
the result does not depend on the magnetic field, this transformation can be
applied for any value of the field but, to compare the zero- and
strong-field values, we choose the same $B$ as in the first line of Eq. (\ref
{sufcon}). Then, 
\begin{equation}
\sigma _{zz}\left( 0\right) =\frac{2e^{2}\tau }{(2\pi )^{3}}eB\int \int \int
d\varepsilon dk_{z}dt_{1}v_{z}^{2}\left( -\frac{\partial f^{0}}{\partial
\varepsilon }\right) .  \label{zero_1}
\end{equation}
Comparing this equation with the first line of Eq. (\ref{sufcon}), we see
that the sufficient condition is equivalent to
\begin{equation}
\int \int \int d\varepsilon dk_{z}dt_{1}\left( -\frac{\partial f^{0}}{%
\partial \varepsilon }\right) \left( v_{z}\langle v_{z}\rangle
-v_{z}^{2}\right) \neq 0.
\end{equation}
Integrating over $t_{1},$ we rewrite the last equation as 
\begin{eqnarray*}
&&\int \int d\varepsilon dk_{z}\left( -\frac{\partial f^{0}}{\partial
\varepsilon }\right) \left( \langle v_{z}^{2}\rangle -\langle v_{z}\rangle
^{2}\right)  \\
&=&\int \int d\varepsilon dk_{z}\left( -\frac{\partial f^{0}}{\partial
\varepsilon }\right) \langle \left( v_{z}-\langle v_{z}\rangle \right)
^{2}\rangle \neq 0.
\end{eqnarray*}
Since the integrand is non-negative, the integral can only vanish if $%
v_{z}=\langle v_{z}\rangle$, which is the case 
if
$v_{z}$ does not depend
on $t_{1}.$ Hence, the sufficient condition is equivalent to the requirement
that 
\begin{equation}
\frac{\partial v_{z}}{\partial t_{1}}=\frac{\partial v_{z}}{\partial k_{x}}%
\frac{\partial k_{x}}{\partial t_{1}}+\frac{\partial v_{z}}{\partial k_{y}}%
\frac{\partial k_{y}}{\partial t_{1}}+\frac{\partial v_{z}}{\partial k_{z}}%
\frac{\partial k_{z}}{\partial t_{1}}\neq 0.
\end{equation}
Recalling that $\mathbf{k}$ satisfies Eq. (\ref{time_1}), we re-write the
last equation as 
\begin{equation}
\left( v_{y}\frac{\partial }{\partial k_{x}}-v_{x}\frac{\partial }{\partial
k_{y}}\right) v_{z}\neq 0
\end{equation}
or, recalling the definition of the operator $\hat{\Omega}$ in Eq. (\ref
{omega}), as 
\begin{equation}
\hat{\Omega}v_{z}\neq 0.
\label{dma}
\end{equation}
Since the sufficient condition (\ref{dma}) coincides with the necessary condition in Eq.~(\ref{omegacondition}), we conclude 
that
Eq.~(\ref{condition}) is both a \emph{%
necessary} and \emph{sufficient} condition for LMR. As a corollary, it also
follows that the strong-field value $\sigma _{zz}\left( \infty \right) $ is
always smaller than or equal to $\sigma _{zz}\left( 0\right) $,
implying that if LMR is finite, it is positive.

Before concluding this section, we would like to comment that our aim was to establish a minimal condition for the appearance of LMR in materials. Specifically, we wanted to explore whether, in the simplest model for scattering, anisotropy of the bandstructure alone can give rise to LMR; the answer turns out to be in the affirmative. It should be pointed out that LMR can also occur due to anisotropic scattering. Indeed, as was shown by Jones and Sondheimer \cite{son2} who chose a special form of the scattering probability to solve the Boltzmann equation exactly, non-zero LMR can occur even for an isotropic spectrum, if the scattering probability is appropriately anisotropic. In general, scattering of Bloch electrons is to be described by a tensor of relaxation times, because different components of momentum relax at different rates. In lieu of a fully microscopic description,
we adopt here an heuristic model, in which the relaxation time, being still a scalar, depends on the point in the $k$ space, 
$\tau =\tau (\mathbf{k})$.
It is easy to see that the necessary condition
for non-zero LMR in this case is modified to: 
\begin{equation}
{\hat\Omega} (\tau v_{z})\neq 0,  \label{newcondition}.
\end{equation}
That means 
that
even if the spectrum alone violates our previous condition (
\ref{condition}),  \emph{i.e.,} ${\hat\Omega} v_{z}=0$, Eq.~(\ref
{newcondition}) may still be satisfied because ${\hat\Omega} \tau $ may be
non-zero. If this is the case, LMR is finite as well.
On the other
hand, an attempt to prove that Eq.~(\ref{newcondition}) is also a sufficient condition in this case fails because of the
following reason. With $\tau =\tau (
\mathbf{k})$, 
expressions for the high-field and zero-field longitudinal conductivities
are still given by Eqs.~(\ref{sufcon}) and (\ref{zero_1}), except that now $\tau$ is inside the integrals.
Following 
same reasoning as before, a sufficient condition for non-zero
LMR would be $\sigma _{zz}(B=\infty )\neq \sigma _{zz}(B=0)$, which now
 implies that $\int \int (\frac{-\partial f^{0}}{\partial \varepsilon }%
)(\langle v^2_{z}\tau \rangle -\langle v_{z}\rangle \langle v_{z}\tau
\rangle )\mathrm{d}\varepsilon \mathrm{d}k_{z}\neq 0$. Unlike the 
previous case, however, the integrand cannot be proven to
be a positive function; therefore,  a non-zero integrand  does not guarantee
that the integral is also non-zero. Therefore, the sufficient condition 
can only be formulated in the integral form, as given above.

\section{\label{sec5}Example: longitudinal magnetoresistance  in graphite}

\begin{figure}[t]
\includegraphics[angle=0,width=0.4\textwidth]{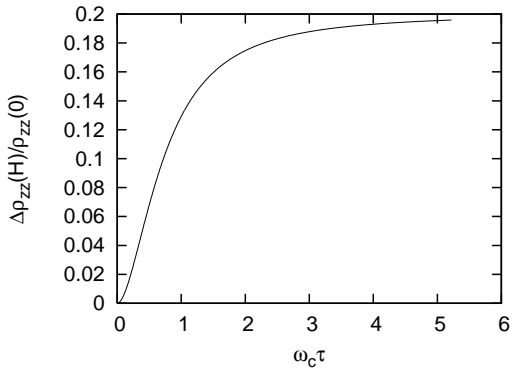}
\caption{Calculated dependence of LMR on magnetic field in graphite. }
\label{fig2}
\end{figure}

As a particular example of a material with significant LMR, we consider the case of graphite, where
a huge- up to three orders of magnitude- LMR effect is observed   when both 
the current and magnetic field are along the {\it c} axis.~\cite{bra}
The crystal crystal structure of graphite
consists of Carbon atoms arranged in hexagonal layers stacked on
top of each other in the Bernal way (ABABAB...). Each unit cell has 4 C
atoms with two inequivalent C atoms in each layer. The resulting Brillouin
zone is a hexagonal prism with very thin elongated FSs along the edges of
the Brillouin zone extended in the direction perpendicular to the plane of
the layers. The energy spectrum of graphite is well described by the Slonczewski Weiss McClure (SWMc)
model \cite{bra} which involves 7 parameters $\gamma_0,...,\gamma_6$,
describing different kinds of interactions between lattice points. 
Here,
$\gamma_0$
and $\gamma_1$ denote in- and out-of-
plane nearest neighbor interactions,  respectively, $\gamma_2,...,\gamma_5$ describe various next-nearest
neighbor interactions, while $\gamma_6$ embodies the difference in the
on-site energies of two inequivalent C atoms in each layer.  Parameter $\gamma_3$ plays a special role as it breaks rotational symmetry of the FS.  
Without $\gamma _{3}$, 
the FS is cylindrically symmetric about the Brillouin zone edge. Therefore, the LMR condition is clearly violated. 
However,  finite $\gamma _{3}$ leads to
``trigonal warping", i.e.,  a three-fold deformation of the FS.
An
expression for energy spectrum of electrons and holes with $\gamma _{3}$ included in
a perturbative way can be written as \cite{mcc} 
\begin{equation}
\varepsilon =\varepsilon _{3}^{0}+A\sigma ^{2}\pm \lbrack B^{2}\sigma
^{4}+2B\gamma _{3}\Gamma \sigma ^{3}\mathrm{cos}(3\alpha )+\gamma
_{3}^{2}\Gamma ^{2}\sigma ^{2}]^{1/2},  \label{spec}
\end{equation}
where $\sigma =\sqrt{3}a_{0}k_{\rho }/2$, $\Gamma =\mathrm{cos}(k_{z}c_{0}/2)$,
 and $\alpha =\pi/2+\phi $, 
with
$a_{0}$ and 
$%
c_{0} $ being the in-plane and out-of-plane lattice constants, respectively.
Also in Eq.~(\ref{spec}), $\varepsilon _{3}^{0}$, $A$ and $B$ are all functions of $k_{z}$ and
contain other interaction parameters. 
Neglecting all the 
next-to-nearest neighbor couplings except for $\gamma_3$ in the spectrum, we have
 $\varepsilon _{3}^{0}=A=0$ and $B=\gamma _{0}^{2}/\gamma _{1}\Gamma $ . 
 With
this approximation, Eq.~(\ref{spec}) can be rewritten [up to $\mathcal{O}%
(\gamma _{3}^{2})$] as : 
\begin{equation}
\varepsilon =\pm \left[ \frac{k_{\rho }^{2}}{2m(k_{z})}+\gamma _{3}\Gamma
\sigma \mathrm{cos}(3\alpha )+\frac{\gamma _{1}\gamma _{3}^{2}\Gamma ^{3}}{%
2\gamma _{0}^{2}}\mathrm{sin}^{2}(3\alpha )\right],  \label{spec2}
\end{equation}
where $m(k_{z})=2\gamma _{1}\Gamma /3a_{0}^{2}\gamma _{0}^{2}$. As
is obvious from Eq.~(\ref{spec}), the terms containing $\alpha $ introduce
the trigonal warping effect in the spectrum. 
Due to the presence of these terms,
the condition for non-zero LMR is satisfied. 
Fig.~\ref{fig2} shows the calculated dependence of LMR on
the magnetic field in units of of $\omega_c\tau$, where $\omega _{c}=3eB/m_{0}$ with
 $m_{0}\equiv m(k_{z}=0)$ in
graphite at zero temperature
 (for $\gamma_0=3.16\;
 \mathrm{eV}, \gamma_1=0.39\; \mathrm{eV,  and }\; \gamma_3=0.315\; \mathrm{eV}$).~\cite{bra}  As expected, 
LMR 
is quadratic
at small fields and eventually saturates at large fields.
However, we note that although this explains qualitatively why graphite has non-zero LMR in the first place,
the curve does not nearly match the experiment quantitatively. Namely,
we find that relative magnetoresistance 
$\Delta \rho _{zz}(B)/\rho _{zz}(0)\equiv \left(\rho_{zz}(B)-\rho_{zz}(0)\right)/\rho_{zz}(0)$
saturates approximately at a value of 0.2.
However, observed value of this ratio is higher by orders of magnitude.\cite{spa} 
This implies that the mechanism of LMR in real graphite
 (as opposed to ideal graphite described by the SWMc model) is 
not simply anisotropy of the FS. 
The disagreement is not
surprising in light of the fact that the mechanism of c-axis transport in graphite 
(not only in finite but also in zero magnetic field) is still not
completely understood and generally believed to be due to 
processes not described by the standard Boltzmann equations, 
e.g., phonon-assisted resonant tunneling through macroscopic defects, e.g., stacking faults,
\cite{ohno,matsubara,gutman}
or disorder-assisted delocalization. \cite{maslov}

\section{\label{sec6}Concluding remarks}

To conclude, we have derived a necessary and sufficient condition that 
an
electronic spectrum should satisfy
in order
to show non-zero longitudinal magnetoresistance within the semiclassical regime of electron transport.
 We find that anisotropy is
essential for non-zero LMR although this anisotropy 
is to be a special kind,
namely, the spectrum must satisfy a particular non-separability condition given by Eq.~(\ref{condition}).
We also show that a phenomenological "damped Bloch electrons" model does not capture essential
physics of semiclassical transport in anisotropic materials. In particular, this model predicts
that LMR is absent not only for isotropic but also anisotropic transport, which is not consistent
either with the predictions of the Boltzmann-equation theory or experiment.

In general, the limiting values of the longitudinal conductivities in the zero- and high-field limits
differ only in how the square of the $z$-component of the electron velocity is averaged over the FS. Excluding
some pathological situations, these two averages can only differ by a numerical coefficient on the order of unity.
Therefore, an LMR effect can, in principle, result from FS anisotropy if its magnitude
does not exceed or comparable to $100$\%.
If, an addition, the lattice structure is such that LMR is only possible only due nearest-neighbor-hopping, 
one should expect even smaller values of LMR. 
In many materials, e.g., copper\cite{pip1} and $\mathrm{Sr}_2\mathrm{RuO}_4$,\cite{hus} the observed
LMR effect is on the order of $10$\%, which is well within the anisotropic-FS mechanism.
However, gigantic LMR effects, such as the one observed in graphite, require explanations  which involve macroscopic 
inhomogeneities of the sample.

\begin{acknowledgements}
We acknowledge stimulating discussions with  
 D. B. Gutman,  A. F. Hebard, S. Hill, P. Kumar, E. I. Rashba, C. Stanton, and S. Tongay.  This
work was supported in part by  NSF-DMR-0908029.
\end{acknowledgements}

\end{document}